\documentstyle[11pt]{article}

\newcommand{\lsim}{
 \mathrel{\setbox0=\hbox{$<$}\raise0.6ex\copy0\kern-\wd0
 \lower0.65ex\hbox{$\sim$}}}

\newcommand{\gsim}{
 \mathrel{\setbox0=\hbox{$>$}\raise0.6ex\copy0\kern-\wd0
 \lower0.65ex\hbox{$\sim$}}}

\begin{document}

\begin{titlepage}
\vspace*{-2cm}
\begin{flushright}
\bf
KFA-IKP(TH)-1995-20\\
TUM/T39-95-17
\end{flushright}

\bigskip

\begin{center}
  {\large\bf Inclusive Heavy-Flavor Production from Nuclei}

  \vspace{2.cm}

  {\large N. N. Nikolaev$^{a,b}$, G. Piller$^{c \,*}$ and
    B. G. Zakharov$^{b}$ }
  \bigskip
  \bigskip

 $^{a)}$Institut  f\"ur Kernphysik,
        Forschungszentrum J\"ulich,\\
        D-52425 J\"ulich, Germany\medskip\\
 $^{b)}$L. D. Landau Institute for Theoretical Physics,
        GSP-1, 117940,\\ ul. Kosygina 2, 117334 Moscow, Russia
        \medskip\\
$^{c)}$ Physik Department, Technische
        Universit\"at M\"unchen,\\ D-85747
        Garching, Germany\\
  \vspace{5.cm}

  {\bf Abstract}
\end{center}
We describe a light-cone wave function formulation for hadroproduction
of heavy-flavors at high energies.
At moderate values of $x_F$ heavy-flavor production can be viewed as
a diffractive excitation of heavy quark-antiquark Fock states, present in
the interacting  gluon from the projectile.
The approach developed here is well suited to address
coherence effects in heavy-quark production from  nuclei at small
values of $x_{t} \lsim  0.1\cdot A^{-1/3}$.

\vspace*{1.3cm}

{\sl \centerline {To be published in Z. Phys. A.}}

\vspace*{1.3cm}

\noindent $^{*}$) Work supported in part by BMBF.

\end{titlepage}

\section{Introduction}

The inclusive production of heavy-flavor quark-antiquark ($Q\bar Q$)  pairs
in high-energy hadron-hadron collisions is one of the standard
applications of perturbative QCD.
Calculations, available in next to leading order
in the strong coupling constant $\alpha_s$,
agree reasonably well with experimental data (for a recent review see
ref.\cite{Fixiea94}).
In the perturbative treatment of heavy-flavor production from
nuclear targets, nuclear modifications of the
total production cross section are usually neglected.
Although this is certainly sufficient on a qualitative level, it is
hard to justify  quantitatively within the framework of perturbative
QCD,
especially since it demands the calculation of higher order Feynman
diagrams which involve different nucleons inside the nucleus.
However the light-cone wave function formalism outlined in this paper
is ideal for an investigation of nuclear effects.
In this formulation heavy-quark production at large energies resembles
diffractive processes.
More insight is gained about the propagation of $Q\bar Q$ pairs
through the nuclear medium.
This matter is closely related  to the effects of ``color
transparency'' and ``color opacity'', which are  discussed
in a large variety of processes, ranging from quasielastic
electron scattering to heavy quarkonium production
(see e.g. \cite{Nikola92,PiMuWe93} and references therein).

Let $s$ be the squared center of mass energy of the collision process.
The Feynman variable $x_F$ is defined as $x_F = 2 P_{Q\bar Q}/\sqrt {s}$,
where $P_{Q\bar Q}$ is the longitudinal momentum of the $Q\bar Q$ pair
in the center of mass of the reaction.
It is common to introduce the variables $x_t$ and $x_b$ via:
\begin{eqnarray}
x_t x_b s &=& M_{Q\bar Q}^2,\\
x_F &=& x_b - x_t\,,
\label{eq:xF}
\end{eqnarray}
with $M_{Q\bar Q}$  being the invariant mass of the produced $Q\bar Q$.
In perturbative QCD  $x_b$ and $x_t$ are identical
to the light-cone momentum fractions of the active beam and target
parton, respectively.
Within the leading order perturbative scheme they enter in the heavy-quark
production cross section as follows:
\begin{equation}
\sigma_{Q\bar{Q}}(s)
=\sum_{i,j}\int dx_{t} dx_{b}\,
f_{t}^{i}(x_{t},\mu^{2})\,
f_{b}^{j}(x_{b},\mu^{2})\,
\sigma_{ij}(x_{t}x_{b}s,\mu^{2})  \, .
\label{eq:pQCD}
\end{equation}
Here $f_{t}^{i}(x_{t},\mu^{2})$ and
$f_{b}^{j}(x_{b},\mu^{2})$ are the
densities of partons "i" and "j", carrying fractions $x_{t}$ and
$x_{b}$ of the light-cone momenta of the colliding target and projectile.
The partonic subprocess $i+j\rightarrow Q+\bar{Q}$
is described by the cross section  $\sigma_{ij}(x_{t}x_{b} s,\mu^{2})$.
It requires  a squared center of mass energy $x_t x_b s  = M_{Q\bar Q}^2
\geq 4 m_Q^2$, where $m_Q$ is the invariant mass of the heavy quark.
The factorization in parton densities and hard partonic subprocesses
is carried out at a typical scale $\mu^2 \sim 4\,m_Q^2$.

In this work we will be concerned with nuclear effects in  inclusive
hadro\-pro\-duction of $Q\bar{Q}$ pairs  which carry a high energy in the
laboratory frame.
In particular  we consider processes at small target light-cone
momentum fractions  $x_t < 0.1$.
Furthermore we will restrict ourselves to moderate $x_F$.
Thus  we concentrate on the kinematic domain where $Q\bar{Q}$ production
is dominated by the gluon fusion subprocess $g+g\rightarrow Q+\bar{Q}$
(see e.g. ref.\cite{PilTho95})
and neither the annihilation of light-quarks nor the
excitation of higher-twist intrinsic heavy-quark components  \cite{VoBrHo92}
are of importance.
The  perturbative QCD cross section (\ref{eq:pQCD})
is then  proportional to the density of gluons in beam and target.
In the case of open-charm production at  typical Tevatron energies
($s \sim 1600\,GeV^2$)
this implies  that we focus on the region $0<x_F<0.5$.

The paper is organized as follows: In Sec.~2 we discuss the
space-time picture of heavy-quark production at small $x_t$, as seen from the
lab frame. Section~3 introduces the light-cone wave function
formulation of heavy-flavor production from free nucleon targets.
This approach is extended to nuclear targets in Sec.~4. Finally
we summarize the main results in Sec.~5.

\section{Lab frame picture of $Q\bar Q$ production at \protect \\
         small $x_t$}

At moderate values of $x_F$
heavy-quark production proceeds via the fusion of a projectile
and target gluon.
In the lab frame, where the target is at rest, the projectile gluon
interacts at small $x_t<0.1$ via quark-antiquark fluctuations present
in its wave function.
In heavy-flavor production only heavy  $Q\bar Q$
Fock states of the incident gluon are of relevance.
Their propagation length  is typically of the order:
\begin{equation}
l_{Q\bar Q} \sim \frac{2\, \nu_G}{4\,m_Q^2} \approx \frac{1}{M\,x_t},
\end{equation}
where $\nu_G = x_b E_{lab}$ denotes the lab frame energy of
the projectile gluon and $E_{lab}$ stands for the beam energy.
For $Q\bar Q$ production from a nuclear target  with mass number $A$
and radius $R_A$  the propagation length $l_{Q\bar Q}$ plays an
important role.
If $l_{Q\bar Q}$ is less then the average nucleon-nucleon
distance in nuclei, $Q\bar Q$ production
takes place incoherently on all nucleons inside the
target.
Consequently then nuclear effects in inclusive heavy-flavor
production will  be absent.
However if the energy of the incident gluon is large, or equivalently
$x_t < 0.1\cdot A^{-1/3}$ is small, the coherence length
will exceed the nuclear size, $l_{Q\bar Q} > R_A$.
Then the formation of $Q\bar Q$ pairs takes place coherently
on the whole nucleus.
This may in principle lead to nuclear modifications
of the  heavy-flavor total production cross section.
\footnote{Note, that this picture is quite similar to deep-inelastic
scattering at small values of the Bjorken variable $x$, leading to nuclear
shadowing (see e.g. \cite{NikZak91,PiRaWe95}).}
The magnitude of these nuclear effects however is controlled by
the transverse size of the $Q\bar Q$ fluctuations and turns out
to be small as discussed below.

Once produced, a  $Q\bar{Q}$ pair  evolves
into hadrons after  a typical formation length
\begin{equation}
l_{f} \sim \frac{2 \pi }{\Delta M_{Q\bar Q}}{\nu_{G} \over M_{Q\bar Q}}
\gg l_{Q\bar{Q}}.
\end{equation}
Here $\Delta M_{Q\bar Q}$ is the characteristic  mass difference of
heavy $Q\bar Q$ states.
Since the formation time $l_f$ is much larger than the coherence
length $l_{Q\bar Q}$, it is justified  to neglect the
intrinsic evolution of a  produced
$Q\bar{Q}$ pair during its propagation through the nucleus.

At large $x_F \sim 1$ the momentum of the projectile  is transfered
collectively to a  heavy $Q\bar Q$ state by several partons in the
beam hadron \cite{VoBrHo92}.
In this case the heavy quark pair  can evidently not be assigned
to a single projectile gluon. Consequently, the excitation
of  heavy-quark components   at $x_F \sim 1$ is quite
different
from the $Q\bar Q$ production mechanism at moderate $x_F$.
In this work we  focus on moderate $x_F$ only, where the above
mechanism is not important.

\section{$Q\bar Q$ production from free nucleons}

As a first step we consider the production of
heavy quark-antiquark pairs through the interaction of
projectile gluons with free nucleons.
The total $Q\bar Q$ production cross section for free nucleon targets
is then obtained by  a convolution with the incident gluon flux.

To leading order in the
QCD coupling constant $\alpha_s$, the incoming projectile gluon
has  a simple Fock state decomposition.
It includes bare gluons, two-gluon states and quark-antiquark
components.
At  small $x_t < 0.1$
the propagation lengths of these Fock
states exceed  the target size
as pointed out in the previous section.
Furthermore, at large projectile energies the transverse
separations and the longitudinal
momenta of all partons in a certain Fock component are
conserved during its interaction with the target \cite{NikZak91}.
Consequently it is most convenient to describe the different
Fock components in terms of  light-cone wave functions in a ``mixed''
representation, given by these conserved quantities.
In this representation the incident, dressed gluon $|G(k_b^2)\rangle$
can be decomposed as:~\footnote{
Wherever it does not lead to confusion we suppress the virtuality
of the incident gluon $k_b^2$.}
\begin{equation}
|G(k_{b}^{2})\rangle = \sqrt{1-n_{Q\bar{Q}}-n_{\xi}}|g\rangle +
\sum_{z,\vec{r}\,} \Psi_{G}(z,\vec{r}\,)|Q\bar{Q};z,\vec{r}\,\rangle
+ \sum_{\xi} \Psi(\xi)|\xi\rangle \, .
\label{eq:Gluon}
\end{equation}
Here $\Psi_{G}(z,\vec{r}\,)$ is the projection of the dressed  gluon wave
function onto a $Q\bar Q$ state  in the  $(z,\vec{r}\,)$-representation,
where  the light-cone variable $z$ represents the fraction of the
gluon momentum carried by the quark and  $\vec{r}$ is the transverse
separation of the $Q$--$\bar{Q}$ in the impact parameter space.
The light quark-antiquark ($q\bar{q}$) and gluon-gluon ($gg$)
Fock components are written as
$\sum_{\xi}\Psi(\xi~)|\xi\rangle$.
Since they are of no importance for our considerations we will omit
them from now on.
The normalization of the $Q\bar Q$ Fock states is given by:
\begin{equation}
n_{Q\bar{Q}}= \int_{0}^{1} dz \int d^{2}\vec{r}\,
|\Psi_{G}(z,\vec{r}\,)|^{2}.
\label{eq:n_QQbar}
\end{equation}
Similarly, $n_{\xi}$ denotes  the normalization of the light-flavor
($q\bar q$ and $gg$) components.
The $Q\bar Q$ wave function $\Psi_G$ of the gluon can be obtained
from the $Q\bar Q$ wave function of a photon $\Psi_{\gamma^*}$,
as derived in ref.\cite{NikZak91}.
The only difference is a color factor and the substitution
of the strong coupling constant for the electromagnetic one:
\begin{eqnarray}
|\Psi_{G}(z,\vec{r}\,)|^{2} &=&
{\alpha_{s}(r) \over 6\,\alpha_{em}}|\Psi_{\gamma^{*}}(z,\vec{r}\,)|^{2}
\nonumber\\
&=&{\alpha_{s}(r)
\over (2\pi)^{2}}\left\{
[z^{2}+(1-z)^{2}]\epsilon^{2}K_1^{2}(\epsilon r) +
m_{Q}^{2}K_{0}^{2}(\epsilon r)\right\}\,.
\label{eq:psi_G}
\end{eqnarray}
$K_{0,1}$ are the modified Bessel functions,
$\alpha_s(r)$ is the running coupling constant in
coordinate-space,
$r = |\vec r\,|$,  and  $\epsilon^{2}=z(1-z)k_b^{2} +m_{Q}^{2}$.

The scattering of an  incident gluon carrying  color $a$ from a
target nucleon at an impact parameter $\vec b$
is described by the scattering matrix $\hat S(\vec b)$:
\begin{equation} \label{eq:SGN}
|G^a,N ;\vec{b}\, \rangle \Longrightarrow
\hat{S}(\vec b\,)|G^a,N;\vec{b}\,\rangle \,.
\end{equation}
In the impact parameter representation
we use  the S-matrix in the eikonal form from ref.\cite{GunSop77}:
\begin{equation}
\hat{S}(\vec b\,)=\exp\left[-i\sum_{i,j}V(\vec{b}+\vec{b}_{i}-\vec{b}_{j})
\hat{T}_{i}^{b}\hat{T}_{j}^{b}\right] \, ,
\label{eq:Smatrix}
\end{equation}
with the one-gluon exchange potential (eikonal function)
\begin{equation}
V(\vec{b}\,)={1 \over \pi}\int d^{2}\vec{k}
\,{\alpha_{s}({\vec k\,}^2)
\,e^{i\vec{k}\cdot\vec{b}} \over \vec{k}^{2} + \mu_{g}^{2}}.
\label{eq:eikonalfunction}
\end{equation}
$\hat{T}_{i,j}^{b}$ are color SU(3) generators
acting on the individual partons  of the projectile and target
at transverse coordinates
$\vec{b}_{i}$ and $\vec{b}_{j}$ respectively.
The effective gluon mass $\mu_g$ is introduced as an infrared
regulator.
In heavy-flavor production $m_{Q} \gg \mu_{g}$,
which ensures that our  final results will
not depend  on the exact choice of $\mu_g$.
Furthermore, as we will show below, $\mu_g$ can be absorbed into the
definition of the target gluon density.
In Eq.(\ref{eq:Smatrix}) and  below
an implicit  summation  over repeated color indices is understood.
In the lowest non-trivial
order the scattering matrix in (\ref{eq:Smatrix})
accounts for the exchange of two t-channel
gluons between the target and projectile.
This has turned out to be successful in describing
high energy hadron-hadron forward scattering processes
\cite{GunSop77}, deep-inelastic scattering at
small values of the Bjorken variable $x$ \cite{NikZak91},
and the diffractive photoproduction of vector mesons \cite{NeNiZa94}.
In Eq.(\ref{eq:SGN}) the leading term is:
\begin{eqnarray}
\left(\hat{S}(\vec b\,)-1\right)&&\hspace{-0.6cm}
|G^a,N;\vec{b}\,\rangle  = \nonumber \\
&&\hspace{-1.2cm}
\frac{-i}{\pi}
\int d^{2}\vec{k}
\,{\alpha_{s}({\vec k\,}^2)
\, e^{i\vec{k}\cdot\vec{b}} \over \vec{k}^{2} + \mu_{g}^{2}}
\left[\sum_j e^{-i\vec k\cdot \vec b_j}
\hat T^b_j \left|N  \right \rangle \right]
\left[\sum_i e^{ i\vec k\cdot \vec b_i}
\hat T^b_i \left| G^a \right\rangle \right]\!.
\label{eq:SG2}
\end{eqnarray}
The last term in Eq.(\ref{eq:SG2}) represents the coupling of an  exchanged
t-channel gluon with  color $b$ to the incident gluon, while the
second to last term describes the coupling to the nucleon target.
The interaction of the incident gluon can be expressed in terms
of its Fock components specified  in Eq.(\ref{eq:Gluon}).
For this purpose, note that
the quark and antiquark of a certain $Q\bar Q$ Fock state with
transverse distance $\vec r$ and light-cone momentum
fraction $z$  are located at impact parameters
$\vec b_{Q} =  \vec r \,(1- z)$ and
$\vec b_{\bar Q} =  -\vec r \,z $
with respect to the parent gluon.
Neglecting light quark-antiquark and gluon-gluon components we obtain
(see Fig. 1):
\begin{eqnarray}
\hspace*{-2cm}\sum_i e^{i\vec k\cdot \vec b_i} \hat T_i^b
&&\hspace{-0.6cm}\left|G^a \right\rangle =
i f_{abc} \sqrt{1-n_{Q\bar Q}}\left| g^c\right\rangle
\nonumber \\
+\sum_{\vec r,z} &&\hspace{-0.6cm}\Psi_G(z,\vec r)
\left\{
\frac{1}{2} \left( e^{i\vec k \cdot \vec r (1-z)} +
                   e^{-i\vec k \cdot \vec r z} \right)
i f_{abc} \left| Q\bar Q_{[8]}^c\right\rangle \right.
\nonumber \\
&&\hspace*{+0.7cm}\left.+\frac{1}{2} \left( e^{i\vec k \cdot \vec r (1-z)} -
                   e^{-i\vec k \cdot \vec r z} \right)
d_{abc} \left|Q\bar Q_{[8]}^c \right\rangle\right.
\nonumber \\
&&\hspace*{+0.7cm}\left.+\frac{1}{\sqrt{6}}
\left( e^{i\vec k \cdot \vec r (1-z)} -
                   e^{-i\vec k \cdot \vec r z} \right)
\delta_{ab} \left|Q\bar Q_{[1]}\right\rangle
\right\}.
\label{eq:finalQQ1}
\end{eqnarray}
As usual $f_{abc}$ and $d_{abc}$  are the
antisymmetric and symmetric SU$(3)$ structure constants.
In terms of the color wave functions of its quark and
antiquark constituents  the color wave function of an octet $Q\bar Q$ pair
with color $c$ is:
\begin{equation}
\left|Q\bar Q_{[8]}^c\right\rangle = \sqrt{2}
\sum_{k,l} \hat T^c_{k l} \left|Q_k\right \rangle
\left|\bar Q_l\right \rangle.
\end{equation}
The indices $k$ and $l$ specify the color of the
quark and antiquark.
Similarly, a  color singlet $Q\bar Q$ state is given by:
\begin{equation}
\left|Q\bar Q_{[1]}\right\rangle = \frac{1}{\sqrt{3}}
\sum_k \left|Q_k\right \rangle  \left|\bar Q_k\right \rangle.
\end{equation}
The scattering state
(\ref{eq:finalQQ1}) can be decomposed
into final state, dressed gluons  and heavy-quark pairs:
\begin{eqnarray}
  \hspace*{-2cm}\sum_i e^{i\vec k\cdot \vec b_i} \hat T_i^b
&&\hspace{-0.6cm}\left|
  G^a\right\rangle = i f_{abc} \left| G^c\right\rangle \nonumber \\
+\sum_{\vec r,z}
&&\hspace{-0.6cm}\Psi_G(z,\vec r) \left\{ \frac{1}{2} \left( e^{i\vec
    k \cdot \vec r (1-z)} + e^{-i\vec k \cdot \vec r z} - 2 \right) i f_{abc}
  \left| Q\bar Q_{[8]}^c\right\rangle \right.  \nonumber \\
&&\hspace*{+0.7cm}\left.+\frac{1}{2} \left( e^{i\vec k \cdot \vec r (1-z)} -
  e^{-i\vec k \cdot \vec r z} \right) d_{abc} \left|Q\bar Q_{[8]}^c
\right\rangle\right.  \nonumber \\ &&\hspace*{+0.7cm}\left.+\frac{1}{\sqrt{6}}
\left( e^{i\vec k \cdot \vec r (1-z)} - e^{-i\vec k \cdot \vec r z} \right)
\delta_{ab} \left|Q\bar Q_{[1]}\right\rangle \right\}.
\label{eq:finalQQ2}
\end{eqnarray}
In this decomposition  the last three terms
represent heavy-quark pairs pro\-du\-ced via the interaction of the incident
gluon with the exchanged t-channel gluon.
According to the derivation leading to Eq.(\ref{eq:finalQQ2})
they have to be seen as
diffractive excitations of heavy-flavor $Q\bar{Q}$ Fock components
present in the wave function of the incident dressed gluon.
The  contribution of these $Q\bar Q$ Fock states to the
heavy-flavor production amplitude disappears if their transverse
size $\vec r$ vanishes, since then
they cannot be resolved by the exchanged t-channel gluon.
This is due to the fact that
interactions of  point-like color octet $Q\bar{Q}$ states are
indistinguishable from interactions of color octet gluons, as
a consequence of color gauge invariance.
Each of the color octet $Q\bar Q$ terms proportional to $f$ and $d$,
as well as the color singlet term in Eq.(\ref{eq:finalQQ2}),
vanishes for $\vec r\rightarrow 0$.

There is a one-to-one correspondence between the
configuration-space derivation of heavy-flavor production,
and the conventional perturbative QCD approach (see Fig.~1).
The excitation of  color singlet $Q\bar{Q}$ states receives
contributions only from $t$- and $u$-channel quark
exchange diagrams (Fig.~1a and Fig.~1b). The same is true for
the excitation of color octet $Q\bar{Q}$ pairs proportional
to the color factor $d$.
The excitation of color octet $Q\bar{Q}$ states
proportional to $f$ receives contributions also from the $s$-channel
gluon exchange process in Fig.~1c.
They yield  the term ``$-2$'' within  the factor
$( e^{i\vec k \cdot \vec r (1-z)} + e^{-i\vec k \cdot \vec r z} - 2)$
in Eq.(\ref{eq:finalQQ2}).
This $s$-channel gluon-exchange contribution  is crucial for the
disappearance  of the corresponding heavy-quark excitation
amplitude at $\vec{r}\rightarrow 0$.

In the following we focus on the diffractively produced $Q\bar Q$
states and omit the gluon contribution in Eq.(\ref{eq:finalQQ2}).
In the light of our discussion in Sec. 2 this should yield the
main contribution to heavy-flavor production at high energies
and moderate $x_F$.
In the lowest non-trivial order we obtain the
heavy-flavor production cross section in gluon-nucleon collisions
by  inserting the $Q\bar{Q}$ component of
Eq.(\ref{eq:finalQQ2}) into Eq.(\ref{eq:SG2}), multiplying
with the complex conjugate,
integrating over the impact parameter space,
summing over the color of the final state and  averaging
over the color of the incident gluon:
\begin{eqnarray}
  \sigma(G N \rightarrow Q\bar Q\, X)
= \frac{2}{3}\int d^2\vec r \int_0^1 d z
  \left|\Psi_G(z,\vec r)\right|^2
\int d^2 \vec k \,
\frac{\alpha_s^2({\vec k\,}^2)\,
{\cal F}(k^2)}{(k^2 + \mu_g)^2}
\nonumber\\
\times \left( 17 - 9 e^{-i \vec k \cdot \vec r z}
- 9 e^{i \vec k \cdot \vec r (1-z)} + e^{i\vec k \cdot \vec r}
\right).
\label{eq:sigmaGN1}
\end{eqnarray}
All relevant information about the target is
in ${\cal F}(k^2)$, which is linked
to the form factor $G_2$
related to the coupling of two gluons to the nucleon:
\begin{equation}
{\cal F}(k^2) = 1 - G_2(k, - k) =
1 - \left\langle N \right| e^{i\vec k\cdot(\vec b_1 - \vec b_2)}
\left| N\right\rangle.
\end{equation}
The vectors $\vec b_1$ and $\vec b_2$ specify the
coordinates of the interacting
quarks in the impact parameter space.
If the size of the target, which carries no
net color, would shrink to zero, the exchanged gluons would decouple
from the target and ${\cal F} = 0$.
In actual calculations we determine  $G_2$ using
a constituent quark wave function for the nucleon of Gaussian
shape, fitted to the electromagnetic charge radius of the nucleon.

The result in Eq.(\ref{eq:sigmaGN1}) can be illustrated
as follows:
Consider a color singlet  gluon-quark-antiquark
$(gQ\bar Q)$ state with a
$Q$-$g$ and $\bar Q$-$g$ se\-pa\-ration $\vec \rho$ and $\vec R$
respectively.
The $Q$-$\bar Q$ separation is then $\vec r = \vec \rho - \vec R$.
Furthermore let $\sigma(r)$ be the cross section for the interaction of
a $Q\bar{Q}$ color dipole of size $r$ with the nucleon target.
In the Born approximation, i.e. with two gluon exchange, one has
\cite{GunSop77,NikZak91}:
\begin{equation}
\sigma(r) = \frac{16}{3} \int d^2\vec k \,\frac{
\alpha_s^2({\vec k\,}^2)
\,{\cal F}(k^2)}
{(k^2+\mu_g^2)^2} \left( 1- e^{i\vec k\cdot \vec r}\right).
\label{eq:dipole}
\end{equation}
Then the $gQ\bar Q$-nucleon cross section
can be written as \cite{NikZak94JETP}
\begin{equation}
\sigma^N_{g Q\bar Q}(r,R,\rho)  = \frac{9}{8}
\left[\sigma(R) + \sigma(\rho) \right] - \frac{1}{8} \sigma(r)\,.
\label{eq:gQQ}
\end{equation}
Comparing with Eq.~(\ref{eq:sigmaGN1}) we find:
\begin{equation}
\sigma(GN\rightarrow Q\bar Q \,X) =
\int d^2 \vec r \int_0^1 d z \,|\Psi_G(z,\vec r)|^2
\sigma^N_{gQ\bar Q} \left( r, - zr, (1-z) r\right).
\label{eq:sigmaGN2}
\end{equation}
Another derivation of this result, based on unitarity, can be found
in ref.\cite{NiPiZa94}.

Let us briefly discuss the properties of $\sigma(GN\rightarrow Q\bar Q \,X)$.
In this respect it is useful to observe that Eq.(\ref{eq:sigmaGN2})
resembles heavy-flavor contributions to  real and virtual
photoproduction \cite{NikZak91,NikZak94JETP}.
The  wave functions corresponding to these processes
are equal, up to normalization
factors (see Eq.(\ref{eq:psi_G})).
While the three-parton cross section (\ref{eq:gQQ})
enters in hadroproduction,
the dipole cross section (\ref{eq:dipole})
is present in photoproduction.
Both cross sections are closely related through
Eq.(\ref{eq:gQQ}).
In the leading  $\log Q^2$ approximation the dipole
cross section (\ref{eq:dipole}) is proportional
to the gluon distribution of the target
\cite{NikZak94JETP,Nikoea93ZP58,Nikoea93InA8}:
\begin{equation}
\sigma(r) \rightarrow \sigma(x_{t},r)={\pi^{2} \over 3}\,r^{2}\,
\alpha_{s}(r)
\left[
x_{t}\,g_t(x_{t},k_{t}^{2} \sim {1 \over r^{2}})\right].
\label{eq:gluon}
\end{equation}
The explicit
dependence on $x_t$  results from higher order Fock
components of the projectile parton, e.g. $Q\bar Q g$ states.
Note that the gluon distribution in Eq.({\ref{eq:gluon}) absorbs
the infrared regularization $\mu_g$. Most important  for our
further discussion is the color transparency property of
the dipole cross section, i.e. its  proportionality to
$r^2$.

{}From the asymptotic properties of the modified Bessel functions
one finds immediately that the squared $Q\bar Q$ wave function
decreases exponentially for $r > 1/\epsilon$.
We therefore conclude  that for
$k_b^{2} \lsim 4m_{Q}^{2}$ small transverse sizes
are relevant for the  $Q\bar Q$ production process:
\begin{equation}
r^{2}\lsim {1\over m_{Q}^{2}} \, .
\label{eq:size}
\end{equation}
For highly virtual gluons, $k_b^{2} \gg 4m_{Q}^{2}$,
the $Q\bar Q$ wave function selects quark pairs with a
transverse size $r^2 \ll 1/4 m_Q^2$.
Their contribution to the production cross section
vanishes like $\sim 1/k_b^2$  due to the color transparency
property of the dipole cross section in Eq.(\ref{eq:gluon}).
Combining Eqs.(\ref{eq:sigmaGN2},\ref{eq:gluon}, \ref{eq:size})
we find:
\begin{equation}
\sigma(GN\rightarrow Q\bar{Q}X) \propto x_{t}\,g_t(x_{t},\mu^{2}
\sim 4m_{Q}^{2})\, .
\label{eq:targetgluon}
\end{equation}
Equation (\ref{eq:targetgluon}) will later be important for a comparison
with the parton model description of heavy-flavor
production in Eq.(\ref{eq:pQCD}).

An important property of
$\sigma(GN \rightarrow Q\bar Q\,X)$ is its infrared stability,
i.e. its convergence in the limit $\vec k \rightarrow 0$.
This is due to the fact that soft t-channel gluons with
$\vec k \rightarrow 0$ cannot resolve the $Q\bar Q$ component of the
incident  beam gluon, and therefore cannot contribute to heavy-flavor
production.

Although $n_{Q\bar Q}$, the normalization of a dressed  gluon
to be found in a $Q\bar Q$ Fock state (\ref{eq:n_QQbar}),
diverges logarithmically  at $r \rightarrow 0$,
the cross section $\sigma(GN \rightarrow Q\bar Q\,X)$ is finite.
This is again implied by color transparency,
since  small size $Q\bar Q$ pairs
cannot be resolved by  interacting t-channel gluons with
wavelengths $\lambda \gsim r$. They are therefore indistinguishable
from bare gluons and cannot be excited into final $Q\bar Q$ states.

We are now in the position to write down the heavy-quark production
cross section  $\sigma_{Q\bar Q}(h,N)$ for hadron-nucleon collisions.
For this purpose we have to
multiply $\sigma(GN\rightarrow Q\bar Q\,X)$ from
Eq.~(\ref{eq:sigmaGN2}) by  the gluon density of the incoming hadron
projectile
and integrate over the virtuality $k_b^2$ of the projectile gluon:
\begin{equation}
{d\sigma_{Q\bar{Q}}(h,N) \over dx_b}=
\int {dk_b^{2} \over k_b^{2}}
{\partial[g_{b}(x_{b},k_b^{2})]
 \over \partial \log k_b^{2}}\,
 \sigma(GN \rightarrow Q\bar{Q}\,X;x_t,k_b^{2}).
\end{equation}
As mentioned above, the leading contributions to
$\sigma(GN\rightarrow Q\bar Q\,X)$ result from the region
$k_b^2 \lsim 4 m_Q^2$.
With the approximation
$\sigma(\!GN\!\rightarrow\!Q\bar Q\,X\!;\!x_t,\!k_b^{2}\!) \approx
\sigma(GN\rightarrow Q\bar Q\,X;x_t,k_b^{2}=0)$ we obtain:
\begin{equation}
{d\sigma_{Q\bar{Q}}(h,N) \over dx_{b}}\approx
g_{b}(x_{b},\mu^{2} = 4m_{Q}^{2})\,
\sigma(GN\rightarrow Q\bar{Q}\,X;x_{t},k_b^{2} = 0) \,.
\label{eq:sigGN}
\end{equation}
Equations (\ref{eq:targetgluon},\ref{eq:sigGN}) describe
the dependence of the production cross section on the
beam and target  gluon densities, in close
correspondence to the conventional
parton model ansatz of Eq.(\ref{eq:pQCD}).
This demonstrates the similarity between the light-cone wave
function  formulation and the conventional parton model approach.
Since  the $g+g\rightarrow Q+\bar{Q}$ cross section
decreases rapidly with the invariant mass of the heavy-quark pair, one finds
dominant contributions to heavy-flavor production for
$x_t x_b s \approx  4 m_Q^2$.
This leads to:
\begin{equation}
{d\sigma_{Q\bar{Q}}(h,N) \over dx_{F}}\approx
\left(1 + \frac{4 m_Q^2}{s x_b^2} \right)
g_{b}(x_{b},4m_{Q}^{2})\,
\sigma(GN\rightarrow Q\bar{Q}\,X;x_{b}-x_{F}) \,,
\label{eq:sighN}
\end{equation}
with $x_b \approx x_F/2 + \sqrt{16 m_Q^2 s + s^2 x_F^2}/2s$.

Let us explore the result (\ref{eq:sighN}) for charm production
in nucleon-nucleon
collisions. To be specific we choose a beam energy
$E_{lab} = 800$\,GeV,
as used in the Fermilab E743 experiment \cite{E74388}.
For the dipole cross section which enters in
Eqs.(\ref{eq:gQQ},\ref{eq:sigmaGN2}) we
employ two different parameterizations.
First we use $\sigma(r)$ from Eq.(\ref{eq:dipole})
with $\mu_{g} = 0.140$\,GeV.
This choice  reproduces measured hadron-nucleon
cross sections at high energies.
Furthermore it has been
successfully applied to  nucleon structure functions at small
$x$ and moderate $Q^2$,
and to nuclear shadowing \cite{NikZak91}.

As an alternative  we
use a parameterization from ref.\cite{NikZak94PLB327}
which includes the effects of higher order Fock states
of the projectile parton.
This choice has been shown to yield a good description of the
small-$x$ proton structure function measured at HERA.
For transverse sizes $r^2 \sim 1/m_c^2$ one finds
in the region $x_{t} \lsim x_{0} = 0.03$ \cite{NikZak94PLB327}:
\begin{equation}
\sigma(x_{t},r) \approx \sigma(r)
\left({x_{0}\over x_{t}}\right)^{\Delta}\, ,
\label{eq:dipoleext}
\end{equation}
with  $\mu_{g}=0.75$\,GeV and $\Delta = 0.4$.
In both parameterizations the squared coupling constant
$\alpha_s^2({\vec k\,}^2)$
appearing in Eq.(\ref{eq:dipole}) has been replaced by
\linebreak
$\alpha_s(r) \,\alpha_s ({\vec k\,}^2)$.

The invariant mass of the charm quark is fixed at $m_c = 1.5$\,GeV.
The gluon distribution of the nucleon projectile enters at
moderate to large values of $x_{b}$. We therefore use the
parameterization of  ref.\cite{Owens91}.
In Fig.2 we compare our results with the
data from the  E743 proton-proton
experiment \cite{E74388}.
In the kinematic region $0<x_F<0.5$,  where our approach is well founded,
both parameterizations of the dipole cross section lead to a
reasonable agreement with the experimental data.




\section{Hadroproduction of $Q\bar{Q}$ pairs from nuclear
\protect \\ targets }


An important aspect of the light-cone wave function formulation
of heavy-flavor production is the factorization of the
$Q\bar Q$ wave function and its interaction cross section
(see  Eq.(\ref{eq:sigmaGN2})).
This feature is crucial for the   generalization
to  nuclear targets.
The derivation of the $Q\bar Q$ production cross section in
(\ref{eq:sigmaGN2}) has been  based upon the observation that
the coherence length $l_{Q\bar Q}$ of a  $Q\bar Q$ fluctuation,
belonging to  the incident gluon, is larger than the target size
for high energies and small $x_t$.
As a consequence the  transverse size  of the heavy-quark pair
is  frozen during the interaction process.
This has led to a diagonalization
of the scattering $\hat{S}$-matrix
in a mixed $(z,\vec{r}\,)$-representation.
In high energy hadron-nucleus collisions
with $l_{Q\bar{Q}} \gsim R_{A}$,
this diagonalization of the $S$-matrix
is also possible.
Consequently we obtain the $Q\bar Q$
production cross section for nuclear
targets by substituting
$\sigma^A_{gQ\bar Q}$ for $\sigma^N_{gQ\bar Q}$ in  Eq.~(\ref{eq:sigmaGN2}).
In the frozen size approximation  the cross section
$\sigma^A_{gQ\bar Q}$  for the scattering of a
 color singlet $gQ\bar Q$ state from
a nucleus with mass number $A$  is given by the
conventional Glauber formalism:
\begin{eqnarray}
\sigma^A_{gQ\bar Q}&=&
2\int d^{2}\vec{b} \left\{
1-\left[1-{1\over 2A}\
\sigma^N_{gQ\bar Q}T(\vec b\,)\right]^{A}\right\} \nonumber\\
&\approx &
2\int d^{2}\vec{b} \left\{
1-\exp\left[-{1\over 2}\
\sigma^N_{gQ\bar Q}T(\vec b\,)\right]\right\}  \, .
\label{eq:siggQQA}
\end{eqnarray}
Here $\vec{b}$ is the impact parameter of  the $gQ\bar{Q}$-nucleus
scattering process,
which must not be confused with the impact parameter of the
$gQ\bar{Q}$-nucleon
interaction in Sec.~3. $T(\vec b\,)$ stands for the optical thickness of the
 nucleus:
\begin{equation}
T(\vec b\,)=\int_{-\infty}^{+\infty} dz \,n_{A}(\vec b,z) \,,
\end{equation}
with the nuclear density ${n_A}(\vec b,z)$ normalized to
$\int d^{3}\vec{r} \,n_{A}(\vec r\,)=A$.
We then obtain:
\begin{equation}
{d\sigma_{Q\bar{Q}}(h,A) \over dx_{b}} \approx
g_{b}(x_{b},\mu^{2} = 4m_{Q}^{2})\,
\sigma(GA\rightarrow Q\bar{Q}\,X;x_t,k_b^{2}=0) \, ,
\label{eq:dsigQQA/dxf}
\end{equation}
where
\begin{eqnarray}
\sigma(GA\rightarrow && \hspace*{-1cm}Q\bar{Q}\,X;x_t,k_b^{2}) =
2 \int d^{2}\vec r \int_0^1 dz \,|\Psi_{G}(z,r)|^{2}
\nonumber\\
&\times& \int d^{2}\vec{b} \left\{
1-\exp\left[-{1\over 2}\
\sigma^N_{gQ\bar Q}T(\vec b\,)\right]\right\}
\equiv
\left\langle \sigma^A_{gQ\bar Q}\right\rangle
\, .
\label{eq:GA}
\end{eqnarray}
In the multiple scattering series (\ref{eq:siggQQA},
\ref{eq:GA})
nuclear coherence effects are controlled by the parameter
\begin{equation}
\tau_{A}=\sigma^N_{gQ\bar Q}\,T(\vec b\,)
\propto \sigma^N_{gQ\bar Q}\,A^{1/3}.
\label{eq:4.5}
\end{equation}
Expanding the exponential in Eq.(\ref{eq:GA})
in powers of $\tau_{A}$ one can identify  terms
proportional to $\tau_{A}^{n}$ which describe contributions to the
total production cross section  resulting from  the coherent
interaction of the $gQ\bar Q$ state with $n$ nucleons inside the
target nucleus.
In leading order, $n = 1$,
 we have the incoherent sum over
the nucleon production cross sections:
\begin{eqnarray}
\sigma(GA\rightarrow Q\bar{Q}\,X) &=&
\int d^{2}\vec{b}\,
T(\vec b\,)\,
\sigma(GN\rightarrow Q\bar{Q}\,X) \nonumber \\
&=&A\,\sigma(GN\rightarrow Q\bar{Q}\,X).
\end{eqnarray}
This is the conventional impulse approximation  component
of the nuclear production cross section, proportional to the
nuclear mass number.
Effects of coherent higher  order interactions are  usually discussed
in terms of the nuclear transparency
\begin{equation}
T_{A}= {\sigma_{Q\bar Q}(h,A) \over {A\, \sigma_{Q\bar Q}(h,N)} } \, .
\label{eq:TA1}
\end{equation}
In the impulse approximation  $T_{A} = 1$. The driving contribution to
nuclear attenuation results  from the coherent interaction of the
three-parton $gQ\bar{Q}$ state with two target nucleons:
\begin{equation}
T_{A}=1-{1\over 4} \,
{\left\langle \left(\sigma^N_{gQ\bar Q}\right)^{2} \right\rangle
\over
\langle \sigma^N_{gQ\bar Q}\rangle}\int d^{2}\vec{b}\,T^{2}(\vec b\,)
\, .
\label{eq:TA2}
\end{equation}
Using Gaussian nuclear densities fitted to the measured
electromagnetic charge radii, $R_{ch} \approx r_0 A^{1/3}\approx
1.1\,fm \,A^{1/3}$, we obtain:
\begin{equation}
T_{A}=1-
\frac{3}{16 \pi r_0^2}
{\left\langle \left(\sigma^N_{gQ\bar Q}\right)^{2} \right\rangle
\over \langle \sigma^N_{gQ\bar Q}\rangle}\,A^{1/3}
\, .
\label{eq:TA3}
\end{equation}
In Fig.3  we present results for the transparency ratio $T_A$
using the two different parameterizations for the
dipole cross section, introduced in the previous section
(see  Eqs.(\ref{eq:dipole},\ref{eq:dipoleext})).
The dipole cross section from Eq.(\ref{eq:dipoleext}) depends
in general on $x_t$, which is however restricted  by the kinematic
constraint $x_t x_b s \geq 4 m_c^2$. We choose $x_t = 0.01$
--- a typical value for current Tevatron energies ($s \sim 1600\,GeV^2$).
For both parameterizations of the dipole cross section we
find only small nuclear effects.
In the commonly used parameterization
$T_A = A^{\alpha -1}$ they translate to $\alpha \sim 0.99$.
This is in agreement with recent experiments at CERN and Fermilab:
In the WA82 experiment \cite{WA8292} with a $340\,GeV$ $\pi^-$ beam one
finds $\alpha = 0.92\pm 0.06$, at an average value of $\bar x_F = 0.24$.
{}From the E769 measurement \cite{E76993}, using a
$250\, GeV$ $\pi^{\pm}$ beam,
$\alpha = 1.00\pm 0.05$ was obtained for $0<x_F<0.5$.

\section{Summary}

We have presented a light-cone wave
function formulation of heavy-flavor production
in high energy hadron-hadron collisions.
At moderate values of $x_F$ we have found that heavy-flavor production
can be viewed as the diffractive excitation of heavy $Q\bar Q$ Fock states
present in the wave function of the interacting projectile gluon.
In the region of applicability ($0<x_F<0.5$) the energy dependence of
recent data on open-charm production in proton-proton collisions is
described reasonably well.
The light-cone wave function formulation is most appropriate to
address the production of heavy-flavors from nuclear targets.
In accordance with recent experiments on open-charm production
we have found small  nuclear effects.
High precision measurements involving heavy nuclear targets
would be necessary to observe any significant nuclear modification of
heavy-flavor production rates.

\bigskip
\bigskip
{\bf Acknowledgments:} G. P. and B. G. Z.
would like to thank J. Speth and the theoretical physics group at
the Institut f\"ur Kernphysik, KFA J\"ulich, for their hospitality
during several visits.
We also thank W. Melnitchouk and W. Weise for a careful reading
of the manuscript.
This work was partially supported by the INTAS grant 93-239 and
grant N9S000 from the International Science Foundation.
\pagebreak\\


\newpage

{\hspace*{-1.35cm}\Large{Figures:}}

\begin{enumerate}
\item[Figure 1:]{
Coupling of the exchanged t-channel gluon
to different Fock components of the incident gluon from the
hadron target. With respect to the gluon-gluon subprocess
(a) and (b) correspond to a $t$- and $u$-channel quark exchange,
and (c) to a $s$-channel gluon exchange process. }

\item[Figure 2:]{The differential cross section
$d\sigma_{c\bar c}/dx_F$ for open
charm production in nucleon-nucleon collisions at
$E_{lab} = 800\,GeV$ calculated via  Eq.(\ref{eq:sighN}).
For the full (dashed)  curve the dipole cross section
from Eq.(\ref{eq:dipole}) \linebreak
(Eq.(\ref{eq:dipoleext})) was used.
The experimental data are from ref.\cite{E74388}.}

\item[Figure 3:]{The $A$ dependence of the nuclear transparency $T_A$.
For the full curve the dipole cross
section from Eq.(\ref{eq:dipole}) was used.
The dashed curve was obtained with the dipole cross
section from Eq.(\ref{eq:dipoleext}) for $x_t = 0.01$.
}

\end{enumerate}

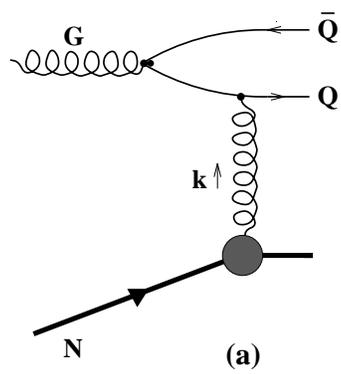
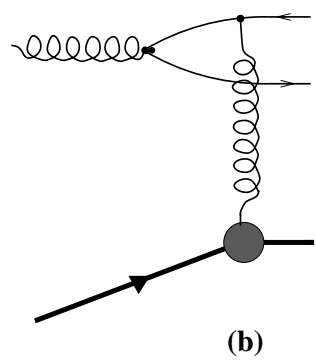
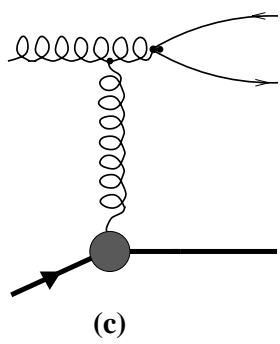
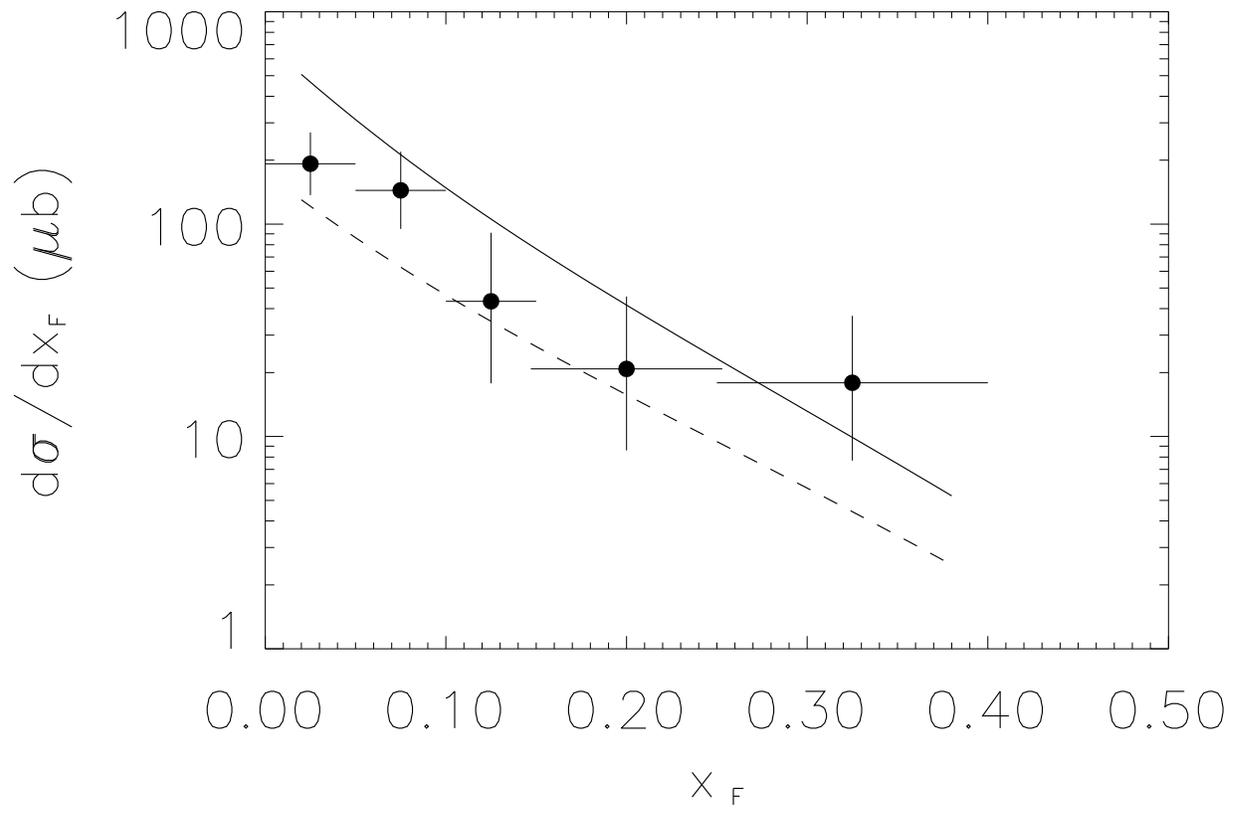
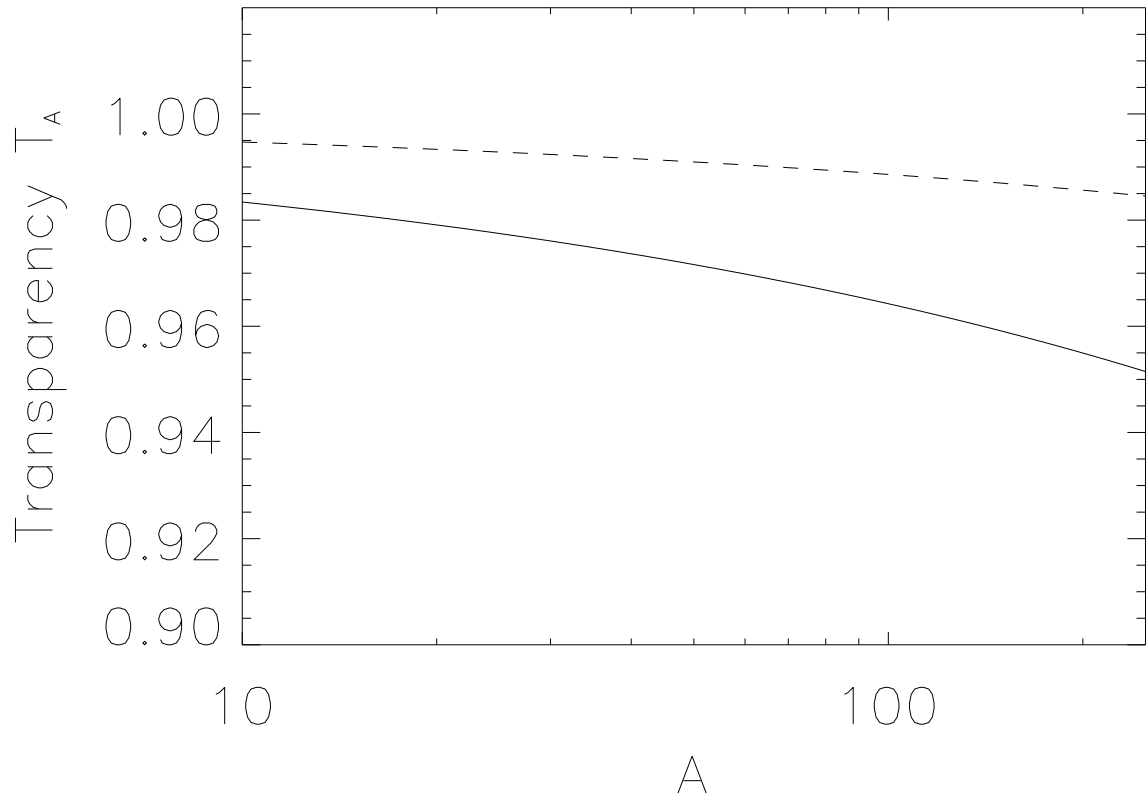
\end{document}